\documentclass[iop]{emulateapj}

\shorttitle{CO diffusion in H$_2$O ice}
\shortauthors{Lauck et al.}

\begin{document}

\title{CO diffusion into amorphous H$_2$O ices}

\author{Trish Lauck}
\affil{Department of Chemistry, University of Virginia,
    Charlottesville, VA 22904, USA}
    
\author{Leendertjan Karssemeijer}
\affil{Theoretical Chemistry, Institute for Molecules and Materials, Radboud University Nijmegen, Heyendaalseweg 135, 6525 AJ Nijmegen, The Netherlands}

\author{Katherine Shulenberger}
\affil{Department of Chemistry, Wellesley College,
    Wellesley, MA 02481, USA}

\author{Mahesh Rajappan}
\affil{Harvard-Smithsonian Center for Astrophysics,
    Cambridge, MA 02138, USA}

\author{Karin I. \"{O}berg}
\affil{Harvard-Smithsonian Center for Astrophysics,
    Cambridge, MA 02138, USA}
\email{koberg@cfa.harvard.edu}

\and

\author{Herma M. Cuppen}
\affil{Theoretical Chemistry, Institute for Molecules and Materials, Radboud University Nijmegen, Heyendaalseweg 135, 6525 AJ Nijmegen, The Netherlands}

\begin{abstract}
The mobility of atoms, molecules and radicals in icy grain mantles regulate ice restructuring, desorption, and chemistry in astrophysical environments. Interstellar ices are dominated by H$_2$O, and  diffusion on external and internal (pore) surfaces of H$_2$O-rich ices is therefore a key process to constrain. This study aims to quantify the diffusion kinetics and barrier of the abundant ice constituent CO into H$_2$O dominated ices at low temperatures (15--23~K), by measuring the mixing rate of initially layered H$_2$O(:CO$_2$)/CO ices.  The mixed fraction of CO as a function of time is determined by monitoring the shape of the infrared CO stretching band.  Mixing is observed at all investigated temperatures on minute time scales, and can be ascribed to CO diffusion in H$_2$O ice pores. The diffusion coefficient and final mixed fraction depend on ice temperature, porosity, thickness and composition. The experiments are analyzed by applying Fick's diffusion equation under the assumption that mixing is due to CO diffusion into an immobile H$_2$O ice. The extracted energy barrier for CO diffusion into amorphous H$_2$O ice is $\sim$160~K. This is effectively a surface diffusion barrier. The derived barrier is low compared to current surface diffusion barriers in use in astrochemical models. Its adoption may significantly change the expected timescales for different ice processes in interstellar environments.

\end{abstract}


\section{Introduction}
In the cold (T$<$20~K) regions of the interstellar medium, the surfaces of dust grains are coated by ice due to a combination of freeze-out of gas-phase molecules and an active grain surface chemistry. Based on observations of ice absorption bands in protostellar and cloud lines-of-sight, the main ice constituent is H$_2$O followed by CO and CO$_2$  \citep[see][for reviews]{Gibb04, Oberg11c}. In most lines of sight there is spectroscopic evidence for separate H$_2$O- and CO-rich ice phases, with CO$_2$ mixed into both \citep{Pontoppidan08}. The diffusion efficiency of molecules between these two phases is poorly constrained, limiting our understanding for how the ice morphology evolves when the grains are heated during star formation. For example, the extent of diffusion of volatile species into H$_2$O ice will regulate the importance of entrapment of volatiles in H$_2$O ice.

Diffusion in ices is also important for the chemical evolution during star formation. Ices are major reservoirs of volatiles and ice chemistry is the proposed pathway to both simple volatiles such as H$_2$O and CH$_3$OH, and to the complex organic molecules observed toward some protostars \citep[e.g.][]{Tielens82,Garrod08,Herbst09}. The efficiencies of these pathways crucially depend on the mobility of the reactants on ice surfaces and in the bulk of the ices \citep{Garrod11,Garrod13a,Vasyunin13}. 

Up until recently there were few laboratory constraints on the diffusion of molecules on top of or inside of interstellar ices \citep{Livingston02,Smith10,Mispelaer13,Karssemeijer14}. Most astrochemical models therefore parameterize surface and bulk diffusion barriers as fractions (30--80\%) of the better understood molecular desorption barriers \citep[e.g.,][]{Tielens82,Garrod11,Chang12}. Surface diffusion is generally modeled as a hopping process between different potential minima with a specific barrier or ensemble of barriers \citep[e.g.,][]{Chang05,Cuppen13}. Within the ice matrix, pore walls provide internal surfaces and diffusion in these pores can be treated similarly to external surface diffusion. `Proper' bulk diffusion has been imagined as either a swapping process \citep{Oberg09e,Fuchs09,Garrod13a}, or a movement into interstitial spaces between molecules \citep[e.g.,][]{Lamberts13,Lamberts14,Chang14}.  Both processes have been modeled to have high barriers compared to surface diffusion.

CO diffusion in H$_2$O ice has been the focus of several studies, both because of its interstellar relevance and because its utility as a model system when evaluating different experimental approaches. \citet{Oberg09e} constrained the segregation rate of thin (10s of monolayers) pre-mixed H$_2$O:CO ices at 23--27~K, under ultra-high vacuum conditions, and obtained a barrier for CO diffusion out of H$_2$O-CO ice mixtures of 300$\pm100$~K (26$\pm9$~meV).  \citet{Mispelaer13} found a diffusion barrier of 120$\pm170$~K (10$\pm15$~meV) for CO diffusion through a thick (100s of monolayers) H$_2$O ice film and into the gas-phase. In a similar set of experiments, \citet{Karssemeijer14} measured the diffusion of CO out of CO:H$_2$O ice mixture, through a thick amorphous ice layer, and into the gas-phase at temperatures of 32--50~K and found a diffusion barrier of 300$\pm170$~K (26$\pm15$~meV). The experimentally determined  barrier of CO diffusion {\it out of} H$_2$O-rich ices of 120--300~K (10--26~meV) is somewhat lower than the $\geq$400~K (34~meV) CO diffusion barrier in current use in astrochemical models \citep{Garrod13a}.  

The experimental results are in some tension with recent simulations of CO diffusion on amorphous H$_2$O, where \citet{Karssemeijer14} found that there are two populations of CO binding sites, one described as a strong-binding nanopore site and the other as a weaker surface site, with diffusion barriers of 80 meV and 30 meV, respectively. Experiments and simulations may be reconcilable if diffusion kinetics in H$_2$O ice mainly depends on the barrier height of the weakest bound CO. To isolate the low-barrier diffusion from diffusion between more strongly bound sites, we study the diffusion kinetics of CO in H$_2$O ice at very low temperatures, where CO in strongly bound sites will be completely immobile. The study of low-temperature diffusion is facilitated by our experimental strategy focusing on CO diffusion {\it into} H$_2$O ice rather than {\it out of} the ice. The latter must be performed at temperatures above the CO desorption temperature of $\sim$30~K.

In summary we aim to quantify the diffusion rate of CO by measuring the mixing rates of initially layered H$_2$O/CO ices. \S2 presents the new experimental set-up designed to explore the physics and chemistry of thin ices, and the experimental procedures and spectral analyses specific for this study. The modeling strategies are explained and motivated in \S3.  \S4  presents the measured mixing kinetics of initially layered H$_2$O/CO ices and their dependencies on ice temperature, thickness, morphology and composition. The experimental results are modeled using a realization of Fick's diffusion equation to extract the CO diffusion barrier. The results are discussed in light of previous experimental and theoretical work, and astrochemical modeling in \S5 and summarized in \S6.
\section{Experimental details \label{exp}}

\subsection{Experimental set-up}

\begin{figure}[htp]
\centering
\includegraphics[width=5.5cm, angle=90]{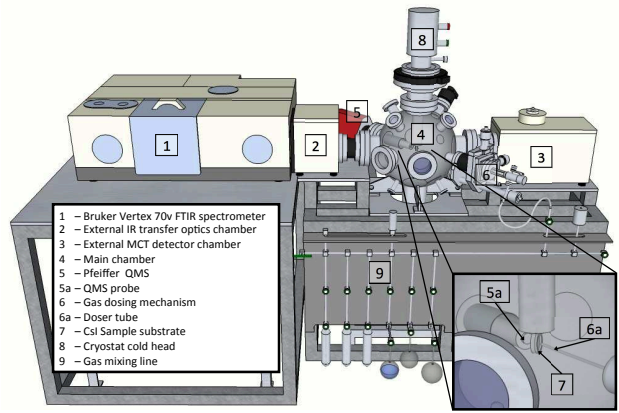}
\caption{Three-dimensional view of the chamber, with the main experimental instrumentation marked. \label{Chamber}} 
\end{figure}

The experiments were all carried out in a new laboratory set-up (Fig. \ref{Chamber}), designed for interstellar ice analog experiments. It consists of a 13" (330.2~mm) spherical stainless steel ultrahigh vacuum (UHV) chamber (custom-made, Pfeiffer Vacuum), evacuated by a Pfeiffer Turbo HiPace 400 pump backed by a DUO 10M rotary vane pump to a base pressure of $\sim$ 10$^{-10}$ mbar at room temperature. The ices are grown on a 2 mm thick IR transparent CsI substrate  with a 19 mm clear view, mounted on an optical ring sample holder at the center of the chamber.  The sample holder is connected to the cold tip of a closed cycle He cryostat (Model CS204B, Advanced Research Systems, Inc.) capable of cooling the CsI substrate down to 11~K. The cryostat is mounted on the top port of the chamber via a differentially pumped UHV rotary seal (Thermionics RNN-400) that allows 360 degree rotation of the CsI substrate inside the chamber without breaking the vacuum. The CsI substrate is mounted onto the nickel plated OHFC copper sample holder using silver gaskets for good thermal contact. A 50 ohm thermofoil heater is installed on the cryocooler tip so that the temperature of the substrate can be varied between 12-350~K. The substrate temperature is regulated by a temperature controller (LakeShore Model 335) using two calibrated silicon diode sensors (accuracy of 0.1~K), one connected directly to the sample holder and the other near the heater element.  

Ice composition, thickness and morphology are monitored through infrared absorption transmission spectroscopy at 4000--400 cm$^{-1}$ using a Bruker Vertex 70v spectrometer with a liquid nitrogen cooled MCT detector.  The IR beam from the spectrometer enters and exits the UHV chamber through KBr ports located on either side of the chamber. The IR beam is focused onto the CsI substrate by a custom designed set of transfer optics (flat and off-axis paraboloidal mirrors) located inside the interferometer chamber, and refocused onto the MCT detector using a similar optics assembly located inside the detector chamber. The spectrometer, the external interferometer chamber and the MCT detector chamber are evacuated to 2 mbar to avoid atmospheric interference with the ice spectra. The spectra reported in this paper have a resolution of 1 cm$^{-1}$ and have been background subtracted. A Pfeiffer quadrupole mass spectrometer (QMG 220M1, mass range 1-100 amu, and resolution of 0.5 amu) is positioned 40 mm off the CsI substrate. It is used to continuously monitor the gas composition during ice growth and warm-up of the deposited ices.  

Ices are grown {\it in situ} by exposing the cold CsI substrate to a constant flow of gas from an independently pumped gas-line (base pressure lower than 10$^{-5}$ mbar) using a gas doser consisting of an xyz stage (enabling positioning 1--90~mm from the substrate window), a precision leak valve (MDC vacuum), a deposition tube (4.8~mm diameter, 14~inches (355.6~mm) long). The deposition tube is connected to a gas-mixing line, which is differentially pumped down to $7 \times 10^{-4}$ mbar, monitored with baratron and pirani gauges. The line contains various ports for glass bulb vessels or stainless steel canisters, which can be interchanged as needed, and several on/off valves, one of which allows the gases to flow to the precision leak valve on the gas doser.

\subsection{Experimental procedure}

The experiments were carried out with CO ($>$99\%, GT \& S, Inc.), CO$_2$ ($>$99\%, Isotec), deionized water and deuterated water (99.9\%, Cambridge Isotope Laboratory, Inc.). The water and deuterated water samples were purified by three freeze-thaw cycles under vacuum. Pure gas samples were prepared by filling one of the bulbs on the gas-mixing line to a pressure of 10~mbar. 10~mbar gas mixtures were made by adding species sequentially to one of the bulbs at the correct proportions, estimated using the pressure gauges and and the known volumes of the different bulbs and tubes within the gas mixing line. This procedure results in maximum $\sim$10\% deviations from target mixtures, based on QMS measurements.

Prior to each experiment the gas flow was set to $5\times10^{-7}$ mbar at room temperature. The CsI window was then cooled down to 12~K, followed by the introduction of a sequence gases or gas mixtures to build up a layered ice structure. The deposition tube was kept close to the surface to minimize porosity in the case of deposition along the surface normal (most experiments) or a highly controlled porosity structure for the handful of ices deposited at an angle. 

\begin{figure*}
\centering
\includegraphics[width=5.5cm, angle=90]{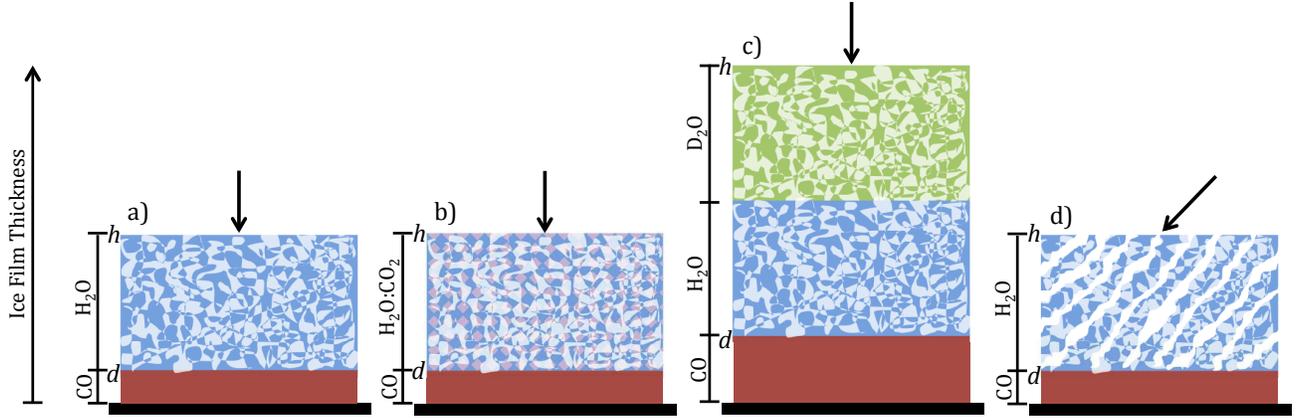}
\caption{An outline of the different ice configurations used on the presented experiments. The top arrows indicate the deposition angle; $d$ indicates the interface of CO and water, and $h$ indicates the vacuum/ice film interface. In most experiments a thin layer of CO ice was covered by H$_2$O ice (a). In a sub-set of experiments the CO layer was instead covered by a H$_2$O:CO$_2$ ice mixture (b). In a third set of experiments the CO was covered by a layer of H$_2$O followed by a third layer of D$_2$O. All these ices were deposited along the surface normal to minimize porosity. A small set of ices were instead deposited at an angle to increase the ice porosity (d). In each case we then monitored the mixing of CO into the ice over layers.
\label{expoverview}}
\end{figure*}  

Figure \ref{expoverview} depicts a general summary of the various experimental sample preparations. To obtain a layered ice structure, CO was deposited first with an ice thickness of 5--60 monolayers (ML). The ice growth was controlled using a known gas flow rate (measured at room temperature) and the deposition time. The achieved CO ice thickness was then measured by acquiring an infrared spectra normal to the sample of the deposited ice using the relationship between column density, integrated optical depth and band strength:

\begin{equation} \label{eq:bandstrength}
N_i= \frac{\int \tau_i(\nu)d\nu}{A_i}, 
\end{equation}

where $N_i$ is the column density (molec/cm$^2$), and $\int \tau_i(\nu)d\nu$ is the integrated optical depth of the IR band area, and $A_i$ is the band strength of the species $i$. The CO ice thickness in ML is calculated using the extracted CO column density and the known CO ice density. The CO ice layer is then covered by a layer of H$_2$O ice (or a H$_2$O ice mixture), whose thickness is regulated and measured using the same procedure. Table \ref{SpecPara} presents the band strengths, band integration range and ice densities of the species used in the different experiments.

	\begin{table}[htp]
	\begin{center}
	\caption{Ice infrared and density data.}
	{\footnotesize
	\label{SpecPara}
	\begin{tabular}{c c c c c c}
	\hline  \hline
 Species  & IR band & Int. range &$A_{i}$&$\rho$\\
  	& & cm$^{-1}$ & cm molec$^{-1}$ & g cm$^{-3}$\\
	\hline	   
  CO & C--O str. &  2120-2170 & 1.1$\times 10^{-17}$\tablenotemark{a}& 0.81\tablenotemark{b}\\
  H$_2$O & O--H str. & 3000-3600 & 2.0$\times 10^{-16}$\tablenotemark{a}& 0.94\tablenotemark{c}\\
  H$_2$O:CO$_2$ & O--H str. &  3000-3600 & 1.6$\times 10^{-16}$\tablenotemark{d}& 0.94\tablenotemark{e}\\
  CO$_2$ & C--O str. &  2310-2370 & 7.6$\times 10^{-17}$\tablenotemark{a} & 1.3\tablenotemark{e}\\
  D$_2$O & O--D str. &  2225-2700 & 1.4$\times 10^{-16}$\tablenotemark{f} & 1.04\tablenotemark{c}\\
  	\hline
  	\end{tabular}
 \\$^{\rm a}${\citet{Gerakines95}}, $^{\rm b}${\citet{Loeffler05}}, $^{\rm c}${\citet{Jenniskens94}}, $^{\rm d}${\citet{Oberg07a}}, $^{\rm e}${\citet{Escribano13}}, $^{\rm f}${\citet{Venyaminov97}}
  }
 	 \end{center}
 	 \end{table}

Following the initial ice characterization the sample was quickly heated (5~K/min) to the targeted experiment temperature and was then maintained at this temperature (15--23~K) for 1.5--4.5 hours while monitoring changes in the ice morphology using the infrared spectral features of CO and H$_2$O. The first IR spectrum was collected immediately upon reaching the target temperature ($\it{t}=0$) and subsequent IR spectra were collected every two minutes with 64 scans per spectrum. During the entire experiment the gas-phase composition was also monitored using the QMS, checking especially for ice desorption -- no CO desorption was observed during any of the experiments. CO desorption was also checked by monitoring the total CO IR band area during each experiment. We find that the total band area is conserved and thus that CO desorption is negligible also at the highest experiment temperature of 23~K. 

\begin{figure*}
\epsscale{0.75}
\plotone{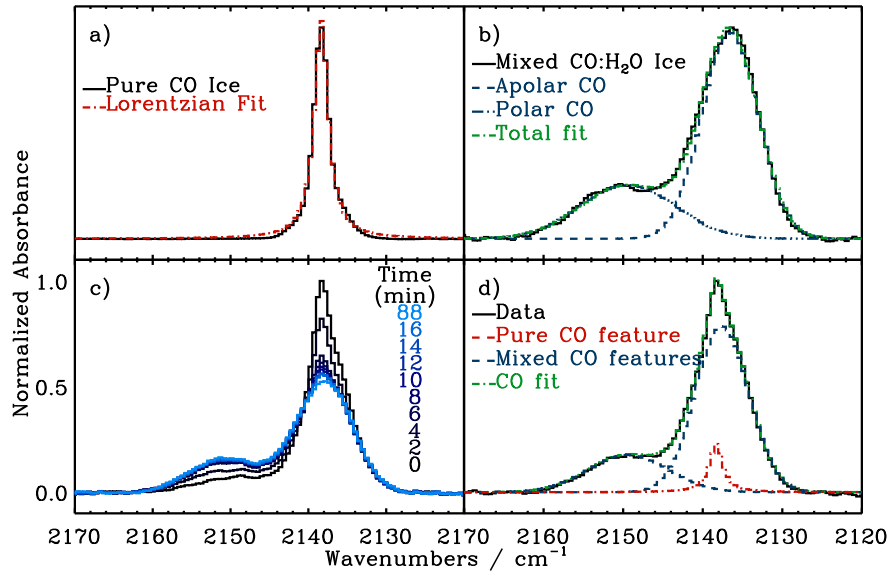}
\caption{Infrared spectra of pure CO ice together with a fitted Lorentzian profile (a), a H$_2$O:CO 5:1 ice mixture (b) together with the spectral Gaussian fits, the changing spectral features as a H$_2$O/CO layered ice mixes (c), and the fit of pure and mixed CO ice components to the end result of (c) using the Lorentzian and Gaussian functions from (a) and (b). 
\label{fits}}
\end{figure*}

\subsection{IR spectral band analysis\label{spectra}}

Pure CO ice and CO mixed with H$_2$O have distinct spectral bands \citep[e.g.,][Fig. \ref{fits}a,b]{Sandford88}.  Pure CO exhibits a Lorentzian IR band profile centered at 2139 cm$^{-1}$ \citep{Bouwman07}, while the spectral feature of CO in a CO:H$_2$O ice mixture can be fit by two overlapping Gaussians, centered at 2137 cm$^{-1}$ and 2152 cm$^{-1}$.  The 2152 cm$^{-1}$ band has been identified with CO molecules interacting with dangling OH sites within the water ice, and is therefore often referred to as the `polar' band. The origin of the 2137 cm$^{-1}$ band is less well understood. It has been attributed to CO interactions with water molecules whose hydrogen atoms are bound within the ice matrix, and also to CO interactions with other CO molecules in H$_2$O ice environment\citep{Schmitt89,Manca01,Al-Halabi04}.

\begin{figure}[htp]
\epsscale{1}	
\plotone{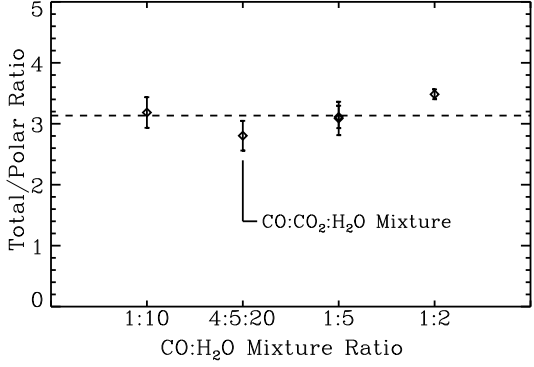}
\caption{The integrated intensity ratio of the total CO band area over the 2152 cm$^{-1}$ (polar) spectral feature in different ice mixtures.
\label{mix}} 
\end{figure}

Figure \ref{fits}c shows the evolving spectra of an initially layered H$_2$O/CO ice kept at 17~K for 90 min. The initial spectra can be fit as a superposition of a pure CO ice spectra and a CO ice mixture spectra, with the pure CO ice dominating. With time the spectra approach the profiles of a completely mixed ice.  The final spectrum (Fig. \ref{fits}d) is well fitted by two Gaussians, and a small Lorentzian. This 3-component fit was automated using the IDL MPFIT function and applied to all experiments and times. Because of the frequency overlap between the pure CO band and the 2137 cm$^{-1}$ mixed ice band, the area of the  2137 cm$^{-1}$ band is less well constrained compared to the 2152 cm$^{-1}$ band.  Since the 2152 cm$^{-1}$ band correlates with mixed CO in water ices, we developed an analysis method using this band without having to isolate the pure band from the 2137 cm$^{-1}$ mixed ice band.  To validate this approach we explored the relative band intensities of the 2137 cm$^{-1}$ and 2152 cm$^{-1}$ bands in five separate CO:H$_2$O ice mixtures using our experimental set-up. The mixtures spanned mixing ratios of 1:2 to 1:10 and also included one experiment with CO$_2$ mixed in. At 12~K all CO mixture spectra present a 1-to-3 intensity ratio between the 2152 cm$^{-1}$ feature and the total CO band (Fig. \ref{mix}). The 2152 cm$^{-1}$ band alone thus provides a good measure of the total amount of CO mixed with H$_2$O. We quantify the amount of mixed CO at each time step in the diffusion experiments with,

\begin{equation} \label{eq:mixed}
N_{\rm mix}= \frac{A_{2152} \times 3}{A_{\rm total}},
\end{equation}
where $N_{\rm mix}$ is the mixed fraction of CO, $A_{2152}$ is the integrated area of the 2152 cm$^{-1}$ feature, and $A_{\rm total}$ is the total integrated area of the CO stretch band.

\section{Diffusion modeling \label{model}}

We use Fick's second law of diffusion to model the observed CO:H$_2$O ice mixing as a function of time. In a one dimensional system, Fick's second law describes the concentration of a diffusing species, $c(z,t)$, as a function of time, $t$, and position, $z$:

\begin{equation} \label{eq:ficks}
\frac{\partial c(z,t)}{\partial t}= D(T)\frac{\partial ^{2}c(z,t)}{\partial z^{2}}.
\end{equation}

This law should apply to the CO:H$_2$O mixing kinetics if the mixing is dominated by the diffusion of the more volatile CO into the H$_2$O matrix described by the temperature dependent diffusion constant $D(T)$. The equation was previously applied to CO diffusion experiments by \citet{Karssemeijer14,Mispelaer13}, where CO was modeled as diffusing through a H$_2$O matrix followed by desorption into the vacuum. 

In the context of Fick's diffusion equation the initially layered H$_2$O/CO ice system is described such that the substrate is at $z=0$,  $z=d$ marks the interface between CO and the H$_2$O layer, and the surface of the H$_2$O layer is at $z=h$.  In this study, the boundary conditions for solving Fick's equation is set by the lack of CO desorption, which entails that the total amount of CO in the system is constant and therefore, the flux of CO at the substrate ($z=0$) and vacuum ($z=h$) interfaces is 0, i.e., $\frac{\partial c(z,t)}{\partial z}=0$ at $z=0$ and $z=h$. In addition the CO concentration was initially assumed to be $c_0$ between $z=0$ and $z=d$, and zero elsewhere. These initial and boundary conditions give the following solution to Equation \ref{eq:ficks}:

\begin{eqnarray} \label{eq:soln}
c(z,t)=\frac{d c_0}{h} + \\ 
\sum_{n = 1}^{\infty}\frac{2 c_0}{n \pi} \sin{\left( \frac{n \pi d}{h} \right)} \cos{\left(\frac{n \pi z}{h} \right)} \exp{\left(- \frac{n^2 \pi^2}{h^2} D t \right)}. \nonumber
\end{eqnarray}

\begin{figure}[htp]
	\epsscale{1}
		\plotone{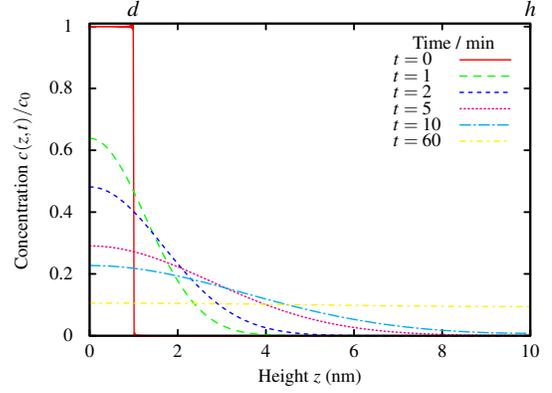}
		\caption{Theoretical progression of CO--H$_2$O through CO diffusion when starting with layered ice films of CO and H$_2$O.  In this example $h=10$~nm, $d=1$~nm, and $D=10^{-16}$ cm$^2$s$^{-1}$.		\label{Conc}} 
		\end{figure}

The generic solution as a function of $z$ and $t$ is shown in Fig. \ref{Conc}. As function of time, CO spreads out into the H$_2$O matrix, approaching the final mixed state, i.e., at $t=0$, all CO is located from 0 to $d$, and with time more and more CO diffuses across the interface $d$, approaching complete mixing at infinite times.

The mixed fraction of CO, $N_{\rm mix}$, is represented as a function of time by integrating Eq. \ref{eq:soln} over the H$_2$O layer and dividing by the total amount of CO: 

\begin{eqnarray} \label{eq:intsoln}
N_{\rm mix}(t) = \frac{1}{d c_0}\int_{d}^{h}c(z,t)dz=\frac{h- d}{h} - \\
\sum_{n = 1}^{\infty} \frac{2 h}{n^2 \pi^2 d} \sin^2{\left( \frac{n \pi d}{h} \right)} \times 
\exp{\left( - \frac{n^2 \pi^2}{h^2} D t \right)}. \nonumber
\end{eqnarray}

Based on Fig. \ref{fits}c, the initial stage of the mixing experiments is not well-described by a complete lack of mixing, however. To allow for mixing on deposition of the H$_2$O ice layer, and mixing during the fast warm-up from the deposition temperature to the mixing temperature at $t<0$ a time offset $t_0$ was added to Eq. \ref{eq:intsoln}. We also added a nuisance parameter $N_0$ to account for experimental uncertainties in the measured ice thicknesses, yielding 

\begin{eqnarray} \label{eq:finsoln}
N_{\rm mix}(t) = N_0\frac{h-d}{h}-\sum_{n = 1}^{\infty} \frac{2 N_0 h}{n^2 \pi^2 d} \sin^2{\left( \frac{n \pi d}{h} \right)} \\
\times 
\exp{\left( - \frac{n^2 \pi^2}{h^2} D (t+t_0) \right)}. \nonumber
\end{eqnarray}

\begin{figure}[htp]
\centering
\includegraphics[width=7.5cm]{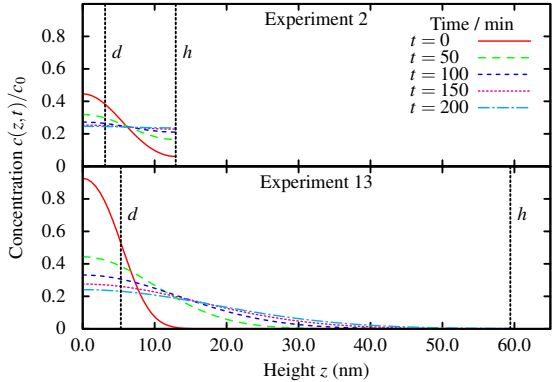}
\caption{The modeled concentration of CO with respect to ice height ($z$), at various time stamps, for thin and thick ice experiments.
\label{Ficks}}
\end{figure}

When using this equation to fit experiments, $h$ and $d$ are taken from spectroscopic measurements, and $D$, $N_0$, and $t_0$ are modeled as free parameters.  Fig. \ref{Ficks} shows the application of this solution to layered H$_2$O/CO experiments of different thicknesses (Exps. 2 and 13 in Table \ref{tbl:results}), demonstrating how the H$_2$O/CO mixing is modeled as a function of time including a measured amount of ice mixing on deposition.  The model solution can be fit to H$_2$O/CO ices with any thickness and mixing ratio, but when the H$_2$O ice is thin compared to the CO:H$_2$O interface, there is no well-defined H$_2$O matrix for the CO to diffuse into, and the derived diffusion coefficients are difficult to interpret. That is, the extracted diffusion coefficients will no longer be applicable for diffusion in pure H$_2$O ice.

.

\section{Results\label{results}}

\begin{table*}[htbp]
\centering
{\scriptsize
\caption{H$_2$O/CO ice mixing experiments ordered based on the main parameter varied within each subset of experiments, together with initial and final fit values and the diffusion rate from Fick's diffusion law analysis in \S3.3. Reported uncertainties do not include systematic ones (see text).}
\label{tbl:results}
\begin{tabular}{c l c c c c c c}
 \hline  \hline
Exp.	&H$_2$O/CO	&$\alpha_{\rm dep}$\tablenotemark{a}	&$T_{\rm mix}$	& $N_{\rm mix}(0)$\tablenotemark{b} & $N_{\rm mix}(t_{\rm final})$\tablenotemark{b} & $D$\\
 &(ML)			&($^{\circ}$)		&(K)			& (\%)		& (\%) & (cm$^2$ s$^{-1}$)\\
 \hline   
 \multicolumn{4}{l} {\it Ice temperature} \\                  
1 & 34/10 & 0 & 12 & 26 & 26 & 3.8$\times 10^{-18}\pm5.8\times 10^{-18}$\\
2 & 31/8 & 0 & 15 & 49 & 64 & 5.2$\times 10^{-17}\pm1.4\times 10^{-18}$\\
3 & 35/9 & 0 & 16 & 51 &  65 & 3.0$\times 10^{-16}\pm1.1\times 10^{-17}$\\
4 & 28/8 & 0 & 17 & 44 &  67 & 3.1$\times 10^{-16}\pm4.6\times 10^{-18}$\\ 
5 & 31/9 & 0 & 20 & 40 &  64 & 2.1$\times 10^{-15}\pm6.5\times 10^{-17}$\\ 
6 & 29/10 & 0 & 23 & 37 &  61 & 2.1$\times 10^{-15}\pm5.1\times 10^{-17}$\\ 
\hline
\multicolumn{4}{l}{\it Ice thickness} \\ 
7 & 13/8 & 0 & 15 & 46 & 55 & 2.1 $\times 10^{-16}\pm1.3\times 10^{-17}$\\
8 & 59/10 & 0 & 15 & 38 & 63 & 1.4 $\times 10^{-16}\pm5.4\times 10^{-19}$\\
9 & 14/9 & 0 & 20 & 53 &  56 & 9.2 $\times 10^{-15}\pm1.3\times 10^{-15}$\\ 
10 & 59/9 & 0 & 20 & 44 & 88 & 2.0 $\times 10^{-15}\pm3.4\times 10^{-16}$\\
11 & 37/5 & 0 & 17 & 68 & 88 & 2.6$\times 10^{-16}\pm4.7\times 10^{-17}$\\ 
12 & 40/14 & 0 & 17 & 11 & 42 & 6.1$\times 10^{-16}\pm1.3\times 10^{-18}$\\ 
\hline
\multicolumn{4}{l}{\it D$_2$O/H$_2$O/CO} \\ 
13 & 89/81/14 & 0 & 15 & 16 & 56 & 1.3$\times 10^{-16}\pm6.1\times 10^{-18}$\\ 
14 & 117/106/30 & 0 & 15 & 8 & 43 & 1.8$\times 10^{-16}\pm2.9\times 10^{-18}$\\
15 & 107/86/59 & 0 & 15 & 5 & 23 & 7.7$\times 10^{-16}\pm1.6\times 10^{-17}$\\
\hline
\multicolumn{4}{l}{\it Porosity\tablenotemark{a}} \\ 
16 & 30/9 & 30 & 15 & 49 & 70 & 8.5$\times 10^{-16}\pm8.9\times 10^{-17}$\\ 
17 & 14/9 & 45 & 15 & 49 & 66 & 6.4$\times 10^{-16}\pm1.4\times 10^{-16}$\\ 
18 & 19/8 & 30 & 20 & 45  & 68 & 3.3$\times 10^{-15}\pm3.9\times 10^{-16}$\\
19 & 26/8 & 45 & 20 & 56 & 81 & 4.0$\times 10^{-15}\pm9.6\times 10^{-18}$\\ 
\hline
\multicolumn{4}{l}{\it Reproducibility} \\ 
20 & 34/10 & 0 & 17 & 26 & 56 & 2.2$\times 10^{-16}\pm4.7\times 10^{-18}$\\ 
21 & 39/10 & 0 & 17 & 30 & 61 & 2.9$\times 10^{-16}\pm4.6\times 10^{-18}$\\ 
22 & 35/10 & 0 & 17 & 32 & 55 & 2.6$\times 10^{-16}\pm2.1\times 10^{-18}$\\ 
\hline
\multicolumn{4}{l}{\it H$_2$O:CO$_2$/CO} \\ 
23\tablenotemark{d} & 46/10 & 0 & 15 & 28 & 48 & 2.0$\times 10^{-16}\pm5.7\times 10^{-18}$\\
24\tablenotemark{d} & 34/9 & 0 & 17 & 29 &  57 & 1.5$\times 10^{-16}\pm2.1\times 10^{-18}$\\ 
25\tablenotemark{d} & 35/10 & 0 & 20 & 24 & 68 & 6.0$\times 10^{-16}\pm5.2\times 10^{-18}$\\ 
\hline
\end{tabular}
\\$^{\rm a}${Porosity of the water layer was varied by changing the angle of incidence, $\alpha_{\rm dep}$, during deposition of the the H$_2$O.}\\
$^{\rm b}${Presented in both the percentage of CO that had mixed and the corresponding monolayers that had mixed.  The uncertainty for each of these values was $<$10\%}.\\
$^{\rm c}${Mixing rate, $k_{\rm diff}$, was determined from $D$ divided by the square of the water (top) layer thickness.}\\
$^{\rm d}${The top layer was a mixture of 4:1 H$_2$O:CO$_2$.}
}
 \end{table*}

\subsection{H$_2$O/CO ice mixing dependencies}

\begin{figure*}
	\epsscale{1}
	\plotone{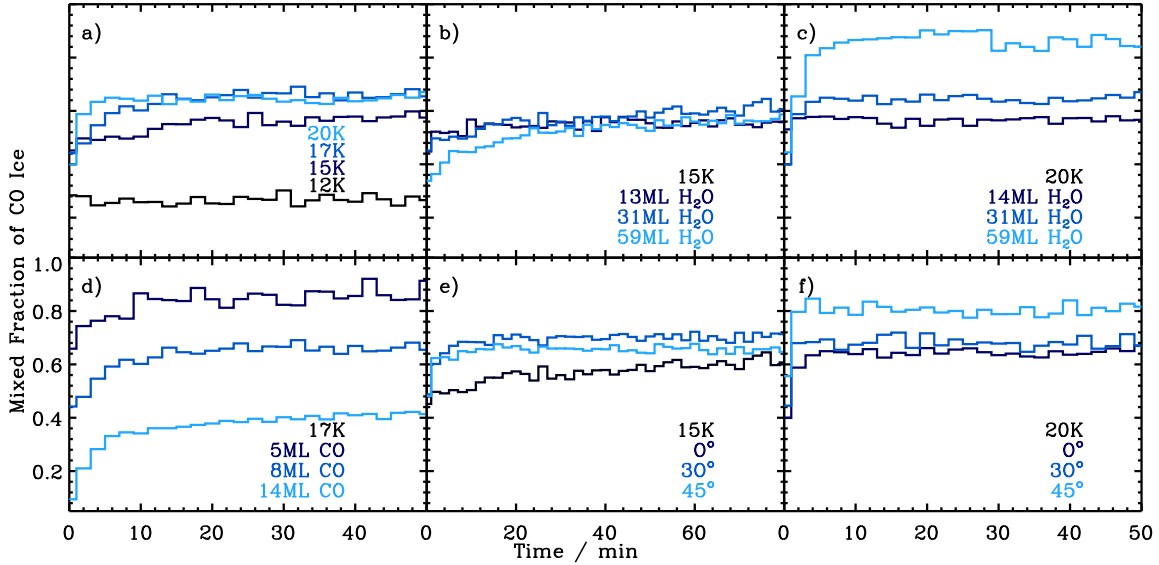}
	\caption{The fraction of mixed CO ice vs. time in initially layered CO/H$_2$O ices.  Panel a) shows the increasing mixing rates with increasing mixing temperature, Exps. 1, 2, 4 \& 5. Panel b) and c) shows the mixing dependencies on H$_2$O ice thickness covering 10~ML of CO at 15 (Exps. 7, 2 \& 8) and 20~K (Exps. 9, 5 \& 10), respectively. Panel d) shows the mixing dependency on CO thickness at 17~K when covered by 35--40~ML of H$_2$O ice, Exps. 11, 4 \& 12. Panel e) and f) show CO mixing with water layers deposited at different angles away from the surface normal; Exps. 2, 16 \& 17 at 15~K, and Exps. 5, 18 \& 19 at  20~K.
	\label{expon}} 
	\end{figure*}
	
Table \ref{tbl:results} lists the experimental details for all initially layered H$_2$O/CO ice mixing experiments. The experiments cover a range of ice mixing temperature (12--23~K), CO and H$_2$O ice thicknesses (5--59~ML and 13--223~ML, respectively), H$_2$O/CO abundance ratios (1.5--14), H$_2$O ice porosities (using different deposition angles), and compositions of the top H$_2$O-rich ice layer.  The fiducial experiment with a H$_2$O/CO ratio of $\sim$4/1 and a mixing temperature of 17~K was repeated four times during the experimental series to test the experimental reproducibility.

Figure \ref{expon} shows the time progression of CO mixing with the H$_2$O ice for a representative subset of the H$_2$O/CO experiments listed in Table \ref{tbl:results}. Above 12~K all layered ice experiments begin to mix within the first few minutes, and the final mixed fraction is typically reached within tens of minutes.  In Fig. \ref{expon}a, the mixing rate, most clearly seen from the initial slope of the curves, increases with temperature, while the final mixed fraction does not. In ice experiments with different ice thickness (Fig. \ref{expon}b--d), the initial mixed CO fraction depends, as expected, on the thickness of the CO under layer. The final mixed CO fraction increases with the thickness of the H$_2$O layer when the CO underlayer thickness is held constant, and decreases with the CO underlayer thickness when the thickness of the H$_2$O ice layer is constant, i.e. it seems to increase with increasing H$_2$O/CO ratio.  At 15~K, the mixing rate is higher for ices deposited at an angle compared to ices deposited along the surface normal (black vs. blue curves in Fig. \ref{expon}e), while at 20~K, the segregation is too fast to visually infer any difference in mixing rate between ices deposited at different angles. Based on previous experiments \citep[e.g.,][]{Kimmel01,Raut07}, an increasing deposition angle results in a more porous H$_2$O ice when the ice is deposited at low temperatures. The experiments thus show that the CO mixing rate increases with ice porosity.

\begin{figure*}[htp]
\centering
\includegraphics[width=17cm,angle=0]{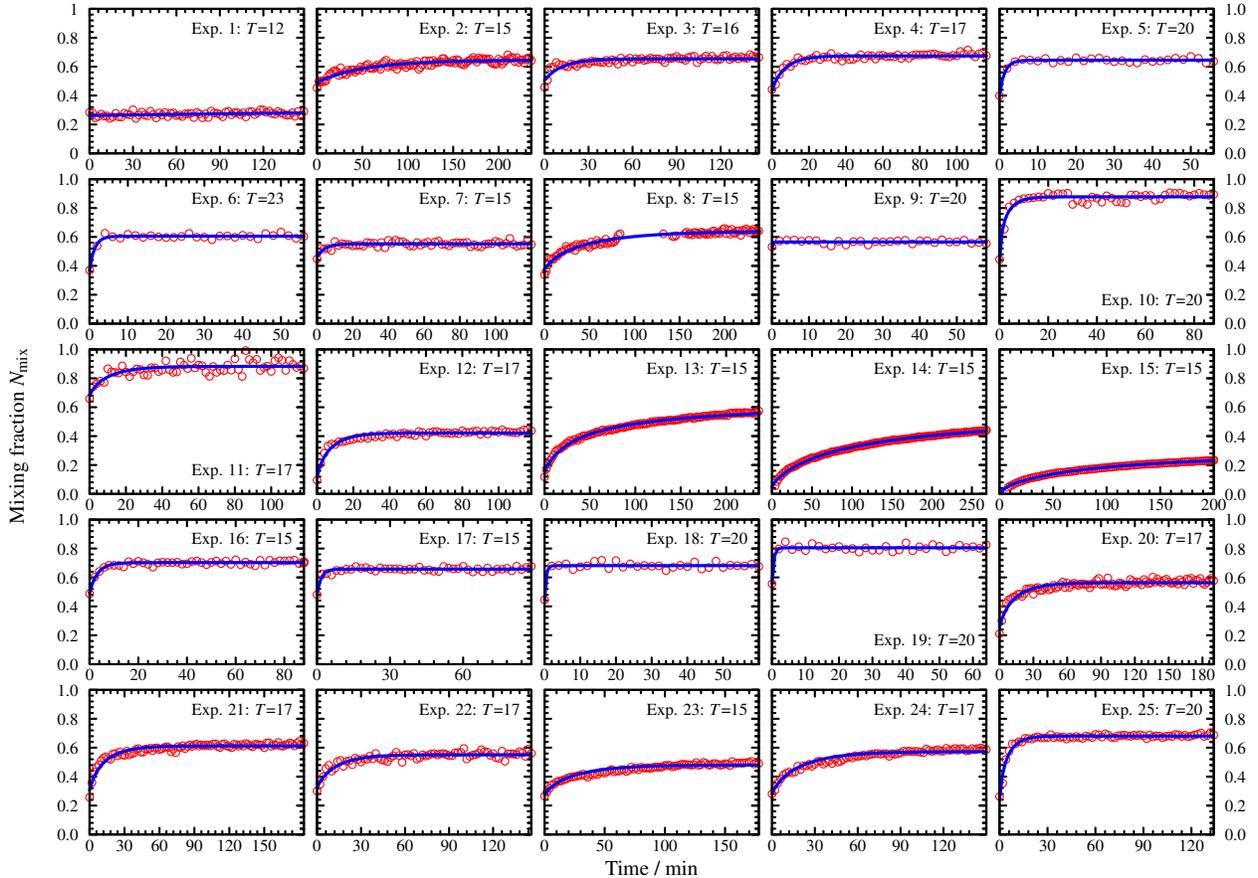}
\caption{Solutions of the Fickian model (solid blue lines) for all experiments (red circles).
\label{FickFits}}
\end{figure*}

To quantify these dependencies we fit all experiments using our solution to Fick's diffusion equation (Eq. \ref{eq:finsoln}). The resulting fits are shown in Fig. \ref{FickFits}, and the extracted initial and final ice mixing data, and the diffusion coefficients are reported in Table \ref{tbl:results}. The fits to the data are generally excellent for the experiments, resulting in small fit uncertainties. As outlined below the experimental+fit uncertainties and the uncertainties in experiments characterized by fast mixing are considerably larger.

There is always a significant amount of mixing at the beginning of each experiment. Between 5 and 68\% of the CO starts out mixed with the H$_2$O (dependent on the CO ice thickness), corresponding to a pre-mixed interface 2--4~ML thick. The mixed CO fraction at the end of each experiment, $N_{\rm mix}(t_{\rm final})$, varies between 26 and 88\%, and increases with increasing H$_2$O/CO ratio as expected from theory: the expected final mixing ratio at $t=\infty$ is $N_0(h-d)/h$. This is best exemplified when comparing experiment 9 and 10, and 11 and 12, and 14 and 15. 

The extracted diffusion coefficients depend strongly on temperature between 12 and 20~K; it increases by three orders of magnitude over this temperature range for the H$_2$O/CO experiments with a thickness of $\sim$32/9 (above 23~K, the diffusion is too fast to effectively measure with our time resolution). For the same ice thickness the fit+experimental uncertainty (1$\sigma$) is estimated to 20\% based on four repeated experiments at 17~K.

The measured diffusion coefficients are expected to be independent off ice thickness as long as diffusion into pure H$_2$O ice dominates and the mixing time scales are long compared to the time resolution. The former condition depends on a combination of ice thickness and H$_2$O/CO ratio. The mixing time scale decreases with increasing temperature and with decreasing ice thickness. The thin ice experiments at high temperatures thus result in the highest model uncertainties. In experiments with short mixing time scales, the extracted diffusion coefficients are lower limits (especially experiments 6--7, 9--10, 18--19). Empirically, measured diffusion coefficients at 15, 17 and 20~K do not depend on ice thickness (within 50\%) as long as the H$_2$O/CO$>$3 and the total ice thickness is $>$36~ML; $cf$ exps. 2, 8, 13 and 14, and exps. 4 and 11. The thickest ices in these comparisons contains a third D$_2$O layer on top of the H$_2$O layer and it is assumed that this isotopic substitution does not affect the CO diffusion into the ice. As inferred visually, the diffusion coefficients do depend on ice porosity. At 15~K, the diffusion coefficient for the ice deposited at an angle (exp. 16) is an order of magnitude higher compared to a the diffusion coefficient derived for a comparable ice deposited along the surface normal (exps. 2 and 8).
 
\begin{figure}[htp]
	\epsscale{	1}
		\plotone{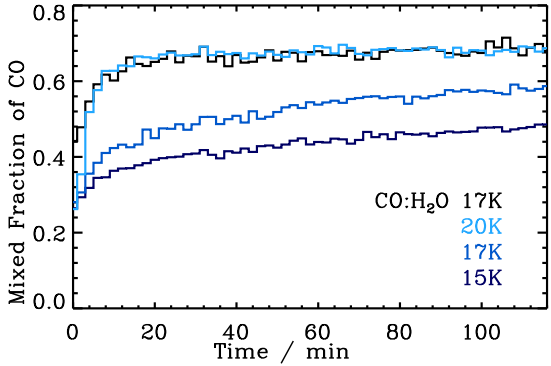}
		\caption{The mixed CO fraction versus time in the CO$_2$:H$_2$O ice mixture experiments (Exps. 23--25).  A 17~K H$_2$O/CO experiment (Exp. 4) is shown for comparison.		\label{CO2mix}} 
		\end{figure}

In addition to the pure ice layer experiments we carried out a few experiments, inspired by interstellar ice compositions, with CO$_2$ mixed into the H$_2$O layer. In these experiments, $\sim$40~ML of a 1:4 CO$_2$:H$_2$O mixture was layered on top of $\sim$10~ML of CO, and the temperature was varied from 15 to 20~K.  While the initial mixed fraction was constant for all three experiments, both the diffusion coefficient and the final mixed fraction increased with temperature, with the coefficient tripling between 15 to 20~K and the final mixed fraction increasing from 0.48 to 0.68. (Fig. \ref{CO2mix}).  The increase in the diffusion coefficient with temperature in these experiments is smaller compared to the H$_2$O/CO mixing experiments without CO$_2$ as a layer component.

\subsection{CO diffusion kinetics and barriers \label{barrier}}

\begin{figure}[htp]
	\epsscale{1}
	\plotone{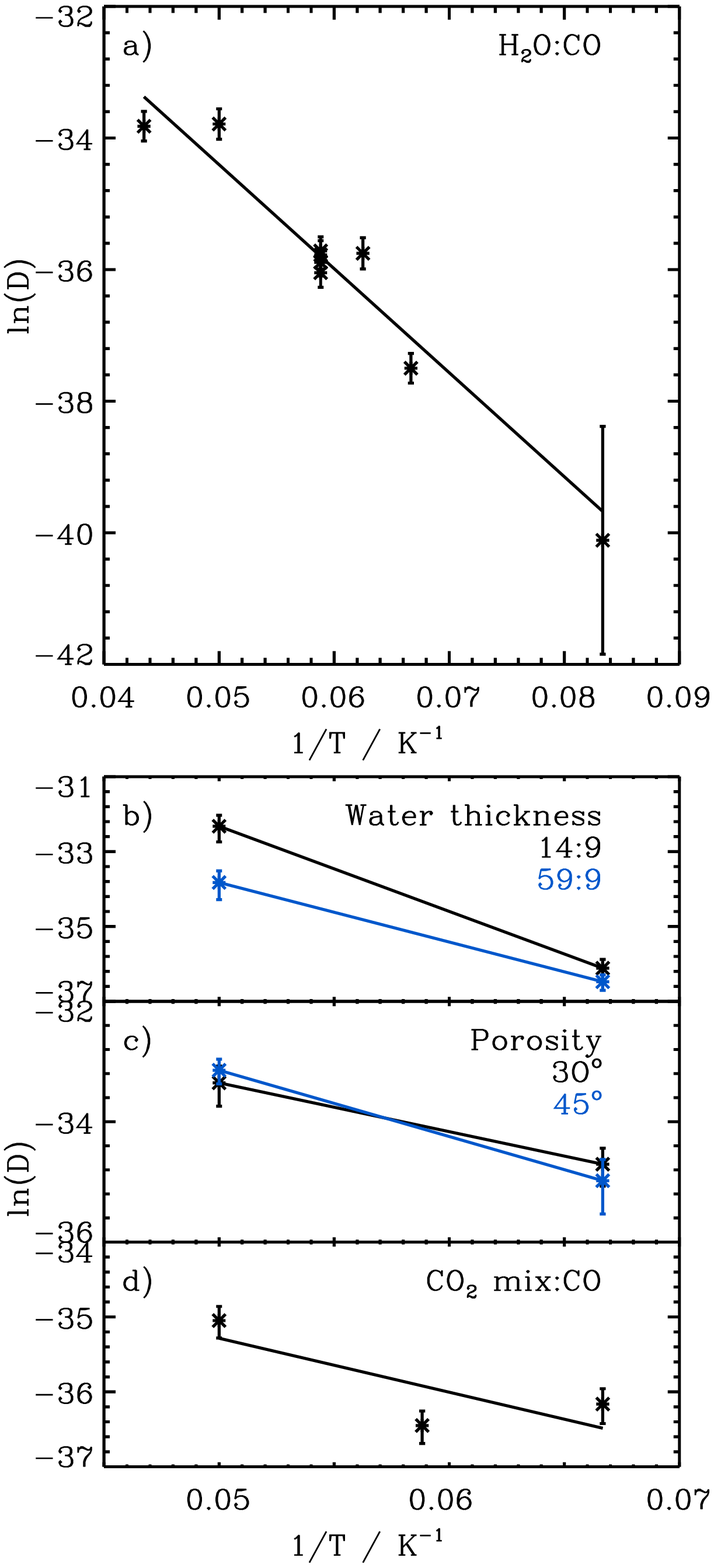}
	\caption{Arrhenius plots incorporating the diffusion coefficients from Exps. 1--6 and 20--22 are plotted in panel a), the diffusion coefficients from the water thickness (Exps. 7 \& 9, 8 \& 10) and porosity (Exps. 16 \& 18, 17 \& 19)  experiments plots in b) and c), and the H$_2$O:CO$_2$ ice mixtures (Exps. 23--25) in panel d).
	\label{arrplot}}
	\end{figure}

We use the diffusion coefficients derived in \S4.1 to constrain the diffusion barrier assuming an Arrhenius type process.  For a single barrier process $D(T)$ should follow the Arrhenius equation, 

\begin{equation} \label{eq:arrhenius}
D=\Gamma \times \textup{e}^{-E_{\rm diff}/T},
\end{equation}

where $\Gamma$ is a pre-exponential factor, $\it E_{\rm diff}$ the diffusion barrier in K, and $T$ the ice temperature. If $D$ is known for multiple ice mixing experiments performed at different temperatures,  $E_{\rm diff}$ is readily extracted by fitting a linear function to $\ln(D)$ versus $1/ T$. The pre-exponential factor can also be extracted from these fits, but its physical meaning is difficult to interpret without additional constraints on the mixing process.  Figure \ref{arrplot}a shows that the diffusion coefficients from  Exps. 1--6, 20--22 (where all parameters except for temperature were kept constant) are well-fit by the Arrhenius equation.  The resulting ice mixing barrier is 158$\pm12$~K.  The reported uncertainties in $D$ in Table \ref{tbl:results} are generally very small because systematic uncertainty is not incorporated.  Based on fits to repeated experiments the experimental uncertainty is closer to $\sim$20\%, and when using the derived diffusion coefficients to calculate barriers, a 20\% uncertainty was therefore added to each data point.

Figure \ref{arrplot}b,c shows the fit of the Arrhenius equation to other subsets of experiments where temperature is the only variable; i.e., `thin' (H$_2$O/CO 14/8~ML) and `thick' (H$_2$O/CO 59/10~ML) experiments, more porous ices defined by the deposition angle, and the experiments with H$_2$O:CO$_2$ ice mixture layered on top. Because of the small number of data points in each case the fit parameters are highly uncertain, but the CO$_2$ mixture experiments and possibly the high porosity experiments do have significantly lower mixing barriers compared to the other experiments (Table \ref{ArrResults}). The pre-exponential factors are also reported. 

\begin{table}[htp]
\begin{center}
{\scriptsize
\caption{Calculated energy barriers and pre exponential factors assuming an Arrhenius-type mixing process.
\label{ArrResults}}
\begin{tabular}{l c c c}
 \hline  \hline
Experiment 	& $\Gamma$ 	&E$_{\rm diff}$ &E$_{\rm diff}$\\
	 		& (cm$^2$ sec$^{-1}$) &(K)  &(meV)\\
 \hline   
Mixing temperature & 3.1 $\times 10^{-12}$ & 158$\pm12$ & 14\\
Thin water layer & 7.9 $\times 10^{-10}$ & 227$\pm26$ & 20\\
Thick water layer & 5.8 $\times 10^{-12}$ & 159$\pm25$ & 14\\
Porosity, 30$^{\circ}$ & 1.9 $\times 10^{-13}$ & 81$\pm27$ & 7\\
Porosity, 45$^{\circ}$ & 9.9 $\times 10^{-13}$ & 110$\pm28$ & 9\\   
CO$_2$ mixture & 1.7 $\times 10^{-14}$ & 72$\pm18$ & 10\\              
\hline
\end{tabular}
}
\end{center}
 \end{table}

\subsection{The CO diffusion mechanism: constraints from H$_2$O ice spectroscopy \label{h2o}}

\begin{figure*}[htpb]
 \epsscale{0.75}
		\plotone{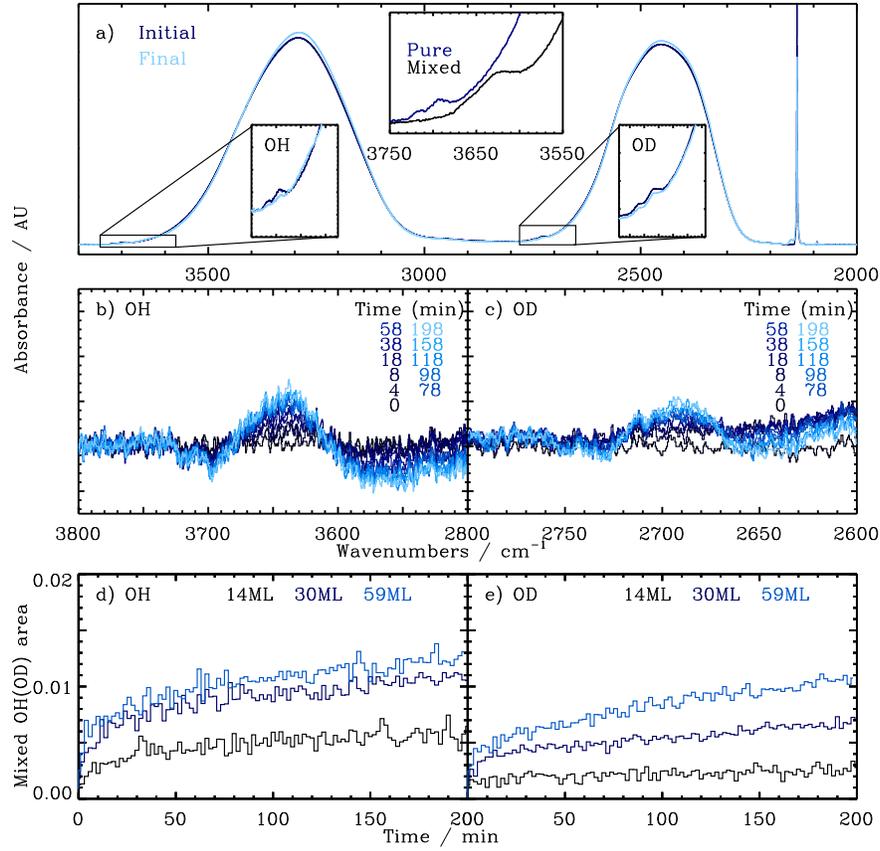}
		\caption{Spectra of pure H$_2$O and a H$_2$O:CO mixture displaying the differences in the dangling OH/OD spectral regions (panel a). Panel b and c show the time evolution of the dangling OH/OD difference spectra in an ice mixing experiment, and panel d and e show the integrated intensity increase with time in the three D$_2$O/H$_2$O/CO experiments (Exps. 13-15).
		\label{DangleOHOD}} 
		\end{figure*}

The H$_2$O ice (and its isotopologue) infrared band profile changes in the presence of CO \citep{Rowland91,Bouwman07}, which can provide independent constraints on the ice mixing rate and mechanism. These changes are most pronounced on the blue wing of the H$_2$O stretching band, where OH-bonds that are not interacting with the hydrogen bonding network in the bulk of the water ice absorb (Fig. \ref{DangleOHOD}a). In pure H$_2$O ice, these dangling OH bonds trace pore surface area, with a double peak IR profile at 3719 and 3697~cm$^{-1}$. H$_2$O molecules that interact with CO present a similar feature slightly shifted to the red \citep{Bouwman07}, at 3639~cm$^{-1}$.  There are corresponding dangling OD bands in pure and mixed D$_2$O ices, with the double pure features at 2750 and 2731~cm$^{-1}$, and the mixed feature at 2692~cm$^{-1}$.

Figure \ref{DangleOHOD}a and b shows that during the diffusion experiments there is a clear loss of the H$_2$O 3719 and 3697~cm$^{-1}$ features and a simultaneous growth of the 3639~cm$^{-1}$ band. Figure \ref{DangleOHOD}d shows the corresponding time series, which displays a similar profile to the mixing time series based on CO spectroscopy, but with a lower signal-to-noise ratio. This increasing exchange of H$_2$O pore spectral features with CO-H$_2$O spectral features indicates that CO diffuses into the H$_2$O ice through pores.

\begin{figure}[htp]
\epsscale{1}
\plotone{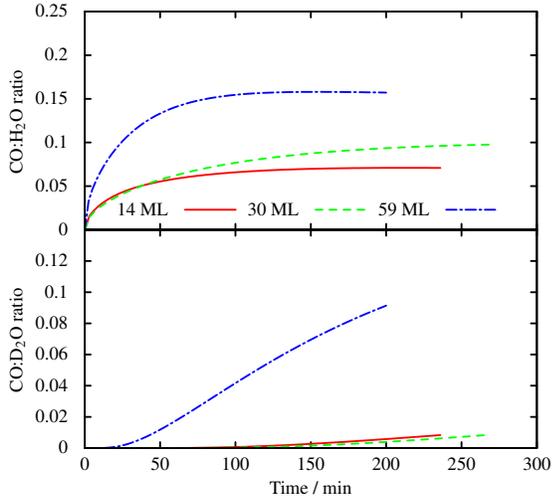}
\caption{The Fickian model of CO diffusion into H$_2$O and D$_2$O when the D$_2$O is layered on top. The model reproduces both the delay into the D$_2$O layer compared to the H$_2$O layer, and how the final portioning of CO between the H$_2$O and D$_2$O layers approaches unity as the CO under layer thickness is increased.
\label{D2Olayerdelay}}
\end{figure}

To constrain the diffusion mechanism further we run several experiments with a D$_2$O layer on top of the H$_2$O layer and simultaneously monitored the changes in the H$_2$O and D$_2$O dangling OH and OD bands. Figure \ref{DangleOHOD}d and e shows that when the bottom CO layer is thin (14~ML), no CO diffusion into the top D$_2$O layer is observed. As the CO ice layer thickness is increased, CO diffusion into the D$_2$O layer is observed, but always at a lower level compare to the intermediary H$_2$O layer. The initial CO mixing rate also appears lower for D$_2$O compared to H$_2$O as would be expected since CO has to travel through H$_2$O before entering into the D$_2$O ice. The low SNR in these experiments as well as an unknown amount of D/H exchange reactions between D$_2$O and H$_2$O prohibits an independent  measure of the diffusion rate based on these experiments.  Figure \ref{D2Olayerdelay} shows, however, that the data qualitatively compares well with the Fickian model solutions for these experiments, based on the fit to the CO spectral time progression modeling n \S4.2.2. CO diffusion into a H$_2$O matrix as described by our solution to Fick's diffusion law thus seems to be a good model of the H$_2$O/CO ice mixing process.

\section{Discussion \label{disc}}

The presented experiments aimed at characterizing CO (surface) diffusion in H$_2$O-rich ices by measuring the mixing of initially layered ices and extracting the diffusion coefficient from the data. This approach should work as long as the observed ice mixing is dominated by CO diffusion into the H$_2$O ice matrix. The general agreement between data and models based on Fick's diffusion law supports that CO diffusion indeed drives the mixing process. Without additional information the CO diffusion could be due either bulk diffusion though e.g., molecular swapping \citep{Oberg09e}, or hopping on the surfaces of nano- and micropores \citep{Karssemeijer14}. Both could {\it a priori} result in the observed diffusion profiles. It is furthermore not necessary that the same diffusion process dominates at all times, since mixing of CO (or other molecules) into the ice could change both the hopping and swapping barriers.

There are several experimental and theoretical considerations that suggest that diffusion of CO in our ice systems is dominated by pore diffusion, however. First, the derived diffusion barrier is very low ($<$170~K). Second, the diffusion rate increases by an order of magnitude at 15~K when the ice porosity is increased. This may be partially due to a reduction in the diffusion barrier (see below), but a more straightforward explanation is that most of the diffusion rate increase is due to an increase in pore area, and that the diffusion process in these experiments proceeds similarly to the low-porosity ice experiments. Finally and perhaps most importantly, the H$_2$O ice spectroscopy in these experiments reveals that as the CO diffuses into the H$_2$O ice, the dangling OH bonds is replaced by spectroscopic features assigned to dangling OH-CO interactions, which is expected if the pore walls become increasingly covered by CO. 

\begin{figure*}[htp]
\centering
\includegraphics[width=3cm, angle=90]{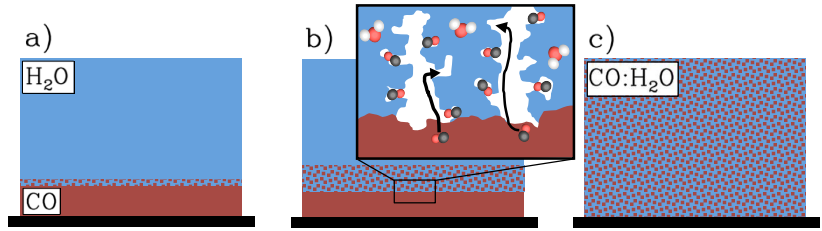}
\caption{Schematic of the mixing process.  The layered system at {\it t} = 0 is shown in a).  Represented in b), mixing occurs, with the inset showing the CO molecules diffusing along the micropore surfaces into the strong-binding nanopore sites.  After some period of time, the layer becomes fully mixed, shown in c).
\label{Schematic}}
\end{figure*}

Simulations provide additional clues on the CO-H$_2$O interactions in these pores. In particular, simulations by \citet{Karssemeijer14} show that water ice surfaces contain multiple sites that interact strongly or weakly with CO. These interactions should manifest themselves in different CO spectral features in CO:H$_2$O ice mixtures, but exact identifications are still lacking. At our low experimental temperatures, CO that becomes bound in a strongly interacting site is expected to be become bound and the diffusion kinetics should be regulated by the number density and the diffusion barrier of the weakly interacting sites. As the strongly bound sites fill up, the CO can diffuse further into the ice before it gets trapped, consistent with the diffusion behavior in the thick D$_2$O/H$_2$O experiments. Spectroscopically, we measure mixing by the growth of CO interacting with dangling OH, which probably correspond to CO that is trapped in a strongly bound site, or nanopore. In this scenario the derived diffusion coefficient thus describes the rate at which a molecule diffuses through the ice along weakly bound sites, while the remaining parameters in the diffusion equation characterizes the availability of empty nanopores/strongly bound sites to diffuse into.  In this scenario, the diffusion length scale of a CO molecule depend on the number of weakly bound sites before a strongly bound site is encountered. The traveled distance should thus increase as more sites become occupied by CO molecules. The maximum diffusion length is limited by the amount of available CO. This is confirmed by the isotopically layered thick ice experiments where it is clear that is is shown that if the CO under layer is thin, no CO will make it all the way through the H$_2$O ice into the D$_2$O ice. Figure \ref{Schematic} shows a schematic of the proposed mixing mechanism of the CO and H$_2$O ices through CO diffusion into the H$_2$O ice via micropores and nanopores.

The experiments further provide evidence for that the already low CO diffusion barrier of $\sim$160~K  is further reduced if the ice porosity is increased and if CO$_2$ is mixed into the H$_2$O ice. The simplest explanation for the low diffusion barrier in the high-porosity experiments is that CO has access to a higher number of low-barrier diffusion pathways. It may also be due to a higher density of  `large' micropores \citep{Kimmel01,Raut07}, i.e., larger than a few molecules across, which could enable CO molecules to ``piggy-back" on top of other CO molecules coating the micropore walls. In other words, the $\sim$100~K barriers derived in the high-porosity ice experiments may reflect the CO-CO diffusion barrier. More experiments are required to distinguish between these scenarios, and indeed to ensure that the extracted small barrier is real. The CO$_2$:H$_2$O ice mixture experiments both result in a lower barrier and in an overall diffusion rate. This suggests that the addition CO$_2$ both changes the overall binding environment, adding lower binding sites compared to pure H$_2$O ice, and reduces the number of pores that CO can diffuse into. 

\subsection{Astrophysical implications}

Based on the presented experiments and analysis, the barrier for CO to diffuse into H$_2$O ice along pore walls is $<$170~K, with a best fit value of $\sim$158~K for a pure, low-porosity ice. This compares well, within the uncertainties, with results from previous studies on CO diffusion in amorphous H$_2$O ice using three different approaches: segregation of CO from CO:H$_2$O ice mixtures \citep{Oberg09e}, CO desorption from CO:H$_2$O ice mixtures \citep{Karssemeijer14}, and CO desorption from layered H$_2$O/CO ices \citep{Mispelaer13}. Summarized in Table \ref{PrevRes}, the derived CO diffusion barriers in these studies all range between 120 and 300~K (10--30 meV).  Considering the good agreement between these very different experiments (including large differences in vacuum and deposition conditions, ice thicknesses and ice morphology), it appears robust that there are CO diffusion barriers that are significantly lower than currently assumed in astrochemical models for either surface or bulk diffusion. 

\begin{table}[htp]
\begin{center}
{\footnotesize
\caption{Summary of our and previous CO diffusion results.
\label{PrevRes}}
\begin{tabular}{l c c c}
 \hline  \hline
 & \multicolumn{2}{c}{Diffusion Barrier}\\
 	&K &meV\\
 \hline  
 \multicolumn{3}{l}{\it Experiments} \\   
Our experiments & 158$\pm12$ &14$\pm1$\\
Segregation\tablenotemark{a}  & 300$\pm100$ & 26$\pm9$\\
Out of thick H$_2$O layer\tablenotemark{b}  & 116$\pm174$ & 10$\pm15$\\ 
Out of CO:H$_2$O mixture\tablenotemark{c} & 302$\pm174$ & 26$\pm15$\\
\hline
\multicolumn{3}{l}{\it Models} \\ 
Gas-Grain\tablenotemark{d} & 400 & 34\\             
CO free on H$_2$O surface\tablenotemark{c} & 348 & 30\\
CO trapped in H$_2$O surface pore\tablenotemark{c} & 929 & 80\\
\hline
\end{tabular}
\\$^{\rm a}${\citet{Oberg09e}}, $^{\rm b}${\citet{Mispelaer13}}, $^{\rm c}${\citet{Karssemeijer14}}, $^{\rm d}${\citet{Garrod13a}}
}
\end{center}
 \end{table}

Our measured CO diffusion barrier describes CO diffusion in H$_2$O ice with pores, i.e. CO diffusion on internal H$_2$O ice surfaces. In non-porous ice, the derived barrier cannot be used to describe ice bulk diffusion, only diffusion on the ice surface.  Furthermore, unless the H$_2$O is impure, resulting in some of the strongly bound sites being occupied by other volatiles, the diffusion length scale with this low barrier may not be very long.  Most astrophysical ices may fall into this category based on the formation process of the H$_2$O ice in the interstellar medium \citep{Garrod13a} and its observed intimate mixing with other volatiles \cite{Pontoppidan08}. The derived barrier thus provides an important constraint on CO surface diffusion; the H$_2$O-CO diffusion barrier of $\sim$160~K is a factor of 2--3 lower than what is currently assumed in astrochemical models for surface diffusion. This discrepancy implies that diffusion of CO, and perhaps many other species, is possible at much lower temperatures than currently assumed.

\section{Conclusions \label{sum}}
The kinetics of CO diffusion into water ice at low temperatures (below the CO desorption temperature) were examined using initially layered ices.  Based on the diffusion dependencies on temperature, ice thickness, porosity, and composition and the subsequent analysis we conclude that:
\begin{enumerate}
	\item CO is mobile in low-porosity amorphous H$_2$O ice at 15--23~K, with a diffusion length scale that depends on the number of available strongly bound sites where CO can become trapped.
	\item The measured CO diffusion into the H$_2$O ice matrix is a pore-mediated process and characterizes CO diffusion on H$_2$O surfaces, rather than bulk diffusion. 
	\item The observed dependence of the CO diffusion on ice temperature implies that the lowest CO diffusion barrier into the pores of amorphous H$_2$O ice  is 160~K .
	\item This barrier is lower in more porous ices and when adding CO$_2$ is added to the H$_2$O matrix.
	\item The derived barrier is low compared to existing values in astrochemical networks, indicative of that surface processes are more efficient than currently assumed.
\end{enumerate}

\acknowledgments
We gratefully acknowledge productive discussion with E. Herbst, D. Pesce, S. Liss, and P. King. D. Graninger assisted with several of the experiments. This work was supported by OT2 funding from Herschel Space Observatory (PI I. Cleeves), European Research Council (ERC-2010-StG, Grant Agreement no. 259510-KISMOL), and the VIDI research program 700.10.427, financed by The Netherlands Organization for Scientific Research (NWO).

\bibliographystyle{aa}

\newcommand{\noop}[1]{}

\end{document}